# Auger Processes in Nanosize Semiconductor Crystals

*Alexander Efros*

*Nanostructure Optics Section*

Naval Research Laboratory, Washington D.C.

## Introduction

In this chapter I discuss the role that various types of Auger processes play in the linear and nonlinear optical properties of semiconductor nanocrystals (NCs). The nanocrystal form of semiconductors was discovered by Ekimov [1], Hengeline [2], and Brus [3] independently in semiconductor doped glasses and colloidal solutions almost 20 years ago. However, rapid progress in semiconductor nanocrystal research, both in chemical synthesis and in physical understanding has been made only in the last ten years. A wide class of semiconductor materials can now be prepared in nanocrystal form including: covalent Si and Ge, III-V compounds (GaAs, GaP, InP), II-VI compounds ( CdSe, CdS, ZnSe, CdTe, PbS and their alloys), and I-VII compounds (CuCl, CuBr, AgBr) (see for review Ref.[4]). Furthermore, in many cases, technology allows one to control the size (from 2 to 30 nm), shapes and surface of these nano-size semiconductor crystals (for an example, see the excellent reviews by Brus [5] and Alivisatos [6]). Realistic calculations of the nanocrystal properties allow us to make a serious analysis of experimental data (for example, see Ref.7). This together with progress in chemical synthesis has brought us to the stage of making different applications of NCs.

Why are nanocrystal quantum dots especially interesting? Because they are a truly new form of matter that can be considered as " artificial atoms." They have linear discrete absorption spectra (like atoms) and photoluminescence tunable over a wide range, from the far infrared to the deep ultraviolet. The nanocrystal size and shape are practically the only two parameters that govern optical transition energies.

On the other hand we can move the NCs around and so form quantum dot molecules, and 3D ordered arrays of these atoms. We can put them in other materials as dopants or join them to a larger molecule to form a super molecule. As a result, instead of 109 elements, we have at our disposal, in principle, an unlimited number of "atomic elements" to form new materials.

These exciting properties lead to numerous potential and actual applications. I mention only two of them that already have interested the industrial world. The first is observation of tunable gain and stimulated emission in CdSe nanocrystal quantum dot solids at room temperature.[8] The second, the successful labeling of a biological molecule by nanocrystals.[9]

Today when we talk seriously about NC applications, the significant role played by Auger processes is especially clear, because they affect all aspects of carrier relaxation and recombination. I will discuss Auger quenching of the photoluminescence (PL), Auger autoionization of the nanocrystals, and Auger-like thermalization of the carrier in the nanocrystals. The impressive experimental data on the PL of the nanocrystal materials will be discussed and analyzed within theoretical models that consider the various Auger processes.

**Quenching of the Nanocrystal Photoluminescence by an Extra Charge**

Extremely impressive evidence of Auger processes is observed in the PL of a single quantum dot. They are clearly seen in the PL of single nanosize quantum dots that show a random intermittence in their PL intensity, an effect that is washed out in ensembles by the contributions from many quantum dots.

The effect was first observed in the PL of a single CdSe quantum dot at room temperature Ref.10 and, since that time, by many other groups at various temperatures [11] and in many other types of nanocrystals: CdS [12], CdTe[13], InP [14] and porous Si [15]. Figure 1 from Ref. [10] shows the time dependence of the photoluminescence intensity of a single CdSe nanocrystal with a radius of 2.1 nm under CW excitation of intensity 0.7 kWcm$^{-2}$.

The time dependence of the photoluminescence intensity exhibits a sequence of "on" and "off" periods. For particles coated by a 7-monolayers-thick ZnS shell the duration of the "on" periods is much longer than that for "bare" particles, which have a very low quantum efficiency they are dark most of the time. Figure 2 from Ref. [10] shows that the duration of the "on" periods decreases with excitation intensity, while that of the "off" periods is almost independent of it.

This unexpected "blinking" behavior of the PL under steady state (CW) excitation condition raises a question about the cause of the "off" periods in the PL intensity. We propose that it is a result of Auger quenching of the photoluminescence in quantum dots that contain an extra charge, i.e., they have become ionized.

Actually, the idea that Auger processes quench PL in a nanocrystal with an extra charge was first proposed to explain the photodarkening effect in semiconductor doped glasses.[16] The decrease in the photoluminescence intensity with time from nanocrystals embedded in glasses under steady state excitation conditions [17] is a general property of all glasses containing nanosize crystals. The effect is reversible: heating the sample above room temperature usually restores the initial PL intensity. We have studied the light-induced photodarkening in glasses doped by CdS nanocrystals. These samples do not show any band edge PL. The PL spectra of these samples are completely dominated by the two transitions from the shallow donor to the two deep acceptors.[18] The inset in Fig. 3 shows a typical PL spectrum of CdS nanocrystals with an average radius of 2.2 nm.

Figure 3 shows the time dependence of the PL intensity of this sample under a steady state excitation condition. We have studied the decrease of luminescence intensity of CdS nanocrystals at a very low excitation intensity. At the maximum intensity we used, $I_{exc}$, the fraction of excited crystals was $10^{-5}$. Figure 3 shows that the rate of the PL degradation as well as the total decrease in the luminescence intensity depends strongly on the exciting intensity. In a first approximation, we characterize this dependence by the time $\tau_i$ at which the intensity of luminescence decreases by factor of two.

The dependence of $\tau_i$ on absorbed power for nanocrystals with different average sizes is shown in Fig.4. One can see that $\tau_i$ decreases rapidly with crystal size and that for all samples the time $\tau_i$ is inversely proportional to the square of absorbed power [16], $P$: $(\tau_I)^{-1} \sim P^2$. This dependence clearly shows that the PL degradation is connected with two photon excitation of the nanocrystals.

Important information about the degradation mechanism was obtained in the thermo-stimulated luminescence (TSL) experiments with these samples.[19] Fig. 5 from Ref. [19,20] compare the TSL spectra of an undoped glass and the glass doped with 3.0 nm radius CdS nanocrystals. Both samples were initially irradiated at $T$=80 K for 1 h. The undoped sample was irradiated by light with energy $\hbar\omega_{ex}$=6eV, which corresponds to interband absorption of the glass, and the TSL was measured within the spectral range of intrinsic luminescence of the glass at $\hbar\omega_{em}$=3.5eV. The TSL curve of the last sample is shown in Fig.5 by the dotted line. It is a typical TSL curve for the well-known $E_1$ capture center of the glass [21]. The doped sample was subjected to light irradiation with $\hbar\omega_{ex}$=3.6eV, which is in the spectral range of interband absorption of nanocrystals, and the TSL spectrum was measured within spectral range of the nanocrystal impurity PL at $\hbar\omega_{em}$=1.7eV. The TSL curve of the doped sample shown in Fig. 5 by the solid line coincides completely with that of the undoped glass. This shows directly that electrons photoexcited in the nanocrystal have a finite probability to leave it and to be captured by the $E_1$ traps in the glass.

Figure 6 shows the spectral dependence of the nanocrystal ionization efficiency, $S_{TSL}$. It is seen that the ionization starts at the photon energy corresponding to the energy gap of the 3.0 nm radius CdS nanocrystals. The increase of the ionization efficiency at $\hbar\omega_b$=4.5 eV corresponds to direct electron excitation over the semiconductor/glass band offset barrier, i.e., to the direct nanocrystal ionization. One can see in Fig. 6 that the electron ejection from the nanocrystal at excitation energies $\hbar\omega < \hbar\omega_b$, which from now on will be called nanocrystal autoionization, is significantly less efficient than the direct crystal ionization. The autoionization of the nanocrystals was also observed at room temperature.

## Autoionization of the Nanocrystals and Auger Quenching of the PL in the Nanocrystal with an Extra Charge

The TSL and the PL time dependence measurements show clearly that the degradation of the PL is determined by the nanocrystal ionization. These experiments suggest the following model that describes light induced PL photodegradation and the single nanocrystal "blinking". The model is shown schematically in Fig. 7.

Optical excitation of an electron-hole pair in a neutral nanocrystal leads to the photon emission as a result of its annihilation. However, it is not always the case for ionized nanocrystals. Excitation of an electron-hole pair in such nanocrystals does not result in PL, because the annihilation energy is taken up not by a photon, but by the extra electron or hole. In a small quantum dot nanocrystal the Auger rate is much larger than the radiative recombination rate, and so the nanradiative Auger recombination quenches all the PL. The nonradiative decay time has been measured in semiconductor doped glasses to be on the order of 10-100 ps [17].

This extra charge can be created in a nanocrystal by optical excitation even if the photon energy is not enough for direct nanocrystal ionization. Two mechanisms of autoionization can be proposed, depending on the size of the nanocrystal, on temperature, on the band offset between the semiconductor nanocrystal and the surrounding matrix, and on the excitation intensity.

The first one is Auger autoionization of a nanocrystal when two electron-hole pairs are excited there simultaneously. In this case the annihilation energy of one of the electron-hole pairs goes into ejecting other electrons from the nanocrystal. The ejected electron (hole) is localized in a trap in the surrounding matrix (for example, at $E_1$ centers of the glass as was found in Ref. [19]) or at the nanocrystal surface.

The thermal or tunnel autoionization of a nanocrystal with a small band offset is another possible mechanism for forming a long-lived ionized state. An optically excited electron or hole in the nanocrystal has a finite probability of thermally exceeding the effective barrier height and then being localized at a deep trap in the matrix or tunneling directly to this trap.

The ejected electron (hole) returns to the quantum dot thermally after a time and restores the "on" periods of the PL. It is the length of time of this return process that determines the duration of the "off" period. The ratio of the "on" to the "off" periods determines the PL quantum efficiency.

The square dependence of the PL degradation rate on the excitation intensity (see Fig. 4) gives a clear indication that two photons are responsible for the nanocrystal ionization. This means that Auger autoionization is the major process leading to the PL degradation in CdS nanocrystals embedded in a glass matrix. Using this assumption we have described the time dependence of the PL degradation in these samples.

On can find four types of the nanocrystals under low excitation intensity of the glass sample: unexcited nanocrystals, nanocrystals with one electron–hole pair, nanocrystals with two electron–hole pairs, and ionized nanocrystals. The system of the rate equations describing the nanocrystal relative population can be written:

$$\begin{aligned}
\frac{dN_0}{dt} &= -W_1 N_0 + \frac{N_1}{\tau_1}; \\
\frac{dN_1}{dt} &= W_1 N_0 - (W_2 + \frac{1}{\tau_1}) N_1 + \frac{N_2}{\tau_2} + \frac{N_+}{\tau}; \\
\frac{dN_2}{dt} &= W_2 N_1 - (\frac{1}{\tau_2} + \frac{1}{\tau_A}) N_2; \\
\frac{dN_+}{dt} &= \frac{N_2}{\tau_A} - \frac{N_+}{\tau};
\end{aligned} \tag{1}$$

where $N_{0,1,2}$ is the nanocrystal concentration with 0, 1, and 2 electron-hole pairs, respectively, $N_+$ is the concentration of ionized nanocrystals, $N = N_0 + N_1 + N_2 + N_+$ is the total nanocrystal concentration, $\tau_{1,2}$ are, respectively, the radiative lifetime of nanocrystals with one and two electron-hole pairs, $1/\tau_A$ is the rate of Auger ionization of a nanocrystal with two electron-hole pairs, $\tau$ is the lifetime for a trapped electron (hole) to return to the nanocrystal, and $W_{1,2} = \sigma_{1,2} I_{exc}/\hbar\omega$, where $\sigma_{1,2}(\omega)$ are the 1 and 2 electron-hole pair excitation cross sections, respectively, of a NC for light of frequency $\omega$, and $I_{exc}$ is the excitation light intensity.

At low excitation intensity the luminescence intensity is proportional to the concentration of nanocrystals containing one electron-hole pair $I_{lum} \sim N_1(t)$. This is because $N_2 \ll N_1$ in our experiment and nanocrystals with two electron-hole pairs do not make a significant contribution to the PL. Therefore the PL degradation is the result of the slow decrease of $N_1$ due to ionization of the crystal. Using Eq. (1) we obtain the time dependence of the relative intensity of the PL, $J_{PL}(t)$:

$$J_{PL}(t) = \frac{N_1(t)}{N_1(0)} = \frac{\tau}{\tau + \tau_i} \exp\left[-t\left(\frac{1}{\tau} + \frac{1}{\tau_i}\right)\right] + \frac{\tau_i}{\tau + \tau_i} , \qquad (2)$$

where $\tau_i$ is the typical time of the nanocrystal Auger autoionization:

$$\frac{1}{\tau_i} = \frac{W_1 W_2 \tau_1}{1 + W_1 \tau_1} \frac{\tau_2}{\tau_2 + \tau_A} . \qquad (3)$$

One can see that at low excitation intensity ($W_1\tau_1 \ll 1$) the rate of the NC autoionization is always proportional to $(I_{exc})^2$ since both $W_1$ and $W_2$ are proportional to $I_{exc}$. The biexciton lifetime $\tau_2 \sim 1$ ns and it is shorter than the time of Auger autoionization $\tau_A$ in practically all sizes of nanocrystals. This allows us to present the crystal ionization rate Eq.(3) in a form:

$$\frac{1}{\tau_i} = \frac{(W_1\tau_1)(W_2\tau_2)}{\tau_A} \qquad (4)$$

that has a very straightforward meaning. In this expression, $(W_1\tau_1)$ is the steady state fraction of the NCs containing a single electron-hole pair, the product $(W_1\tau_1)(W_2\tau_2)$ is the steady state fraction of the crystals containing two electron-hole pairs, and $1/\tau_A$ is the probability of the Auger ionization of these crystals.

The time dependence $J_{PL}(t)$ completely describes the experimental data presented in Fig. 3, and allows us to measure the dependence of nanocrystal ionization rate $\tau_i$ as a function of the NC radius (see Fig. 8) and the time of the nanocrystal neutralization. The nanocrystal ionization rate $\tau_i$ in Fig. 8 increases strongly with the decrease of the nanocrystal radius. However, the time of the nanocrystal neutralization $\tau$ measured in our experiment at nitrogen temperature is within the experimental error of 25% independent of the crystal size and excitation intensity and turns out to be $\tau_{77K}=(1.3\pm0.25)$ h. From the size dependence of $\tau_i$ we have also calculated the size dependence of Auger ionization time $\tau_A$ with use of Eq. 4. The dependence (see Fig. 9) shows that $\tau_A$ decreases from 100 ns to 1 ns with nanocrystal size.

The reversible PL degradation observed in CdS nanocrystals is a very common property of different nanocrystal ensembles. It has been observed in porous Si at helium temperature [22]. Figure 10 shows the PL spectra of porous Si at helium temperature excited by light with $\omega_{exc}$=2.8 eV before and after degradation by a 20 W/cm$^2$ laser at an energy of 2.54 eV for 20 min. One can see that the PL intensity decreases considerably after sample illumination. Heating the sample up to room temperature and cooling it back to 5 K restores the initial intensity of the PL. Analysis of these measurements within the above model of Auger autoionization and Auger quenching of the PL [22] gives us $\tau_A$ = 17 ns and the time of the crystal neutralization at helium temperature $\tau_{5K}$= 3 h. One can see that the time of Auger ionization measured in Si nanocrystals is consistent with those times measured in CdS nanocrystals.

All these results suggest that the random intermittence of the PL of a single NC and the light induced PL photodegradation of NC ensembles reflect essentially the same process: nanocrystal ionization and the quenching of the PL in auto-ionized nanocrystals. The only difference is the time in which it takes an electron or hole to return to the NC. For the carrier localized at a trap in the glass or at some surface states, the time should exponentially depend on the temperature $\tau \sim exp(\Delta E/kT)$, where $T$ is the temperature and $\Delta E$ is the depth of the trap. In the case of photodegradation (low temperature and deep traps), this time is hours and in the random intermittence of the single dot PL (room temperatures and shallow traps) this time is seconds. However, by decreasing the temperature one can significantly increase the time of electron (hole) return to the nanocrystal and transfer the random intermittence into photodegradation.

## Rate of Auger Processes in Nanocrystals

In the model proposed above the rate of Auger processes is much faster than the rate of the carrier radiative recombination. This fact is also supported by experimental measurements of the time of the nonradiative Auger recombination, which was found to be on the order of 10-100 ps [17], while the time of the carrier radiative recombination is always on the order of several nanoseconds. The high rate of nonradiative Auger recombination is a surprising effect because in bulk wide-gap semiconductors such as CdS or CdSe the rate of Auger processes is considerably suppressed. In homogeneous bulk material both energy and momentum are conserved in the Auger process. This leads to a kinematic energy threshold for the reaction and a recombination rate that depends exponentially on the ratio *Eg/T* of the semiconductor energy gap *Eg* to the temperature *T*. Therefore, Auger processes are considerably suppressed in wide-band semiconductors.

There are two reasons why Auger processes are so efficient in the NC. First of all, the momentum is not a good quantum number for electron and hole motion in semiconductor nanocrystals. The second is an enhancement of Coulomb interactions between the carriers confined in a small volume of the NC. The rate of Auger processes in nanocrystals can be calculated using Fermi's golden rule:

$$\frac{1}{\tau_A} = \frac{2\pi}{\hbar} \sum_{k,l,m} |\langle \Psi^i | v(\vec{r}_1, \vec{r}_2) | \Psi^f_{k,l,m} \rangle|^2 \, \delta(E_i - E_f) \qquad (5)$$

where $E_i$, $E_f$ and $\Psi^i$, $\Psi^f$ are the total energies and wave functions of the initial and final multi-electron state of the quantum dot, respectively. The sum goes over all final states of the system $(k,l,m)$ and $v(\vec{r}_1,\vec{r}_2) = e^2/(\kappa|\vec{r}_1 - \vec{r}_2|)$ is the Coulomb potential, where $\kappa$ is a dielectric constant.

Accurate calculation of the matrix element $M = \langle \Psi^i | v | \Psi^f \rangle$ in Eq. 5 is the major problem in Auger rate estimation. One can surmise that this matrix element evaluated for wave functions confined within the NC should be on the order of $M \sim e^2/\kappa a$, where $a$ is the crystal radius. However, this estimation does not take into account the very large momentum that the Auger electron has in the final state. This momentum is $k_f \sim \sqrt{2m_e E_g}/\hbar$, where $m_e$ is the electron effective mass, because practically the whole annihilation energy of the electron-hole pair ($\sim E_g$) is transferred into the electron kinetic energy. The electron wave function oscillates very rapidly inside the nanocrystal because of $k_f >> 1/a$. That drastically decreases the value of the matrix element $M$. The integrand in $M$ can be written as a product of a rapidly oscillating function and a smooth function, $f(r)$. Integration over the internal volume of the nanocrystal gives essentially zero contribution to the integral. However, such integrals can be expanded as a power series in the small parameter $(a\, k_f)^{-1} << 1$, the coefficients of which are determined by the poles of the smooth function [23].

In the case of semiconductor nanocrystals the most important pole is at the nanocrystal surface [16]:

$$M = \langle f(\vec{r}) \exp(-i\vec{k}\,\vec{r}) \rangle \approx f'|_a\, (ak_f)^{-2} + f''|_a\, (ak_f)^{-3} + f'''|_a\, (ak_f)^{-4} \qquad (6)$$

where the coefficients of $(a\ k_f)^{-2}$ and $(a\ k_f)^{-3}$ vanish because of continuity of the wave function and its derivative at the crystal surface. The first nonvanishing term is proportional to $f'''|_a$. A significant contribution to the matrix element $M$ comes only from the surface of the nanocrystal. However, this matrix element does not vanish for any electron-hole pair state because, unlike the bulk, there is no momentum conservation in the nanocrystals that leads to a temperature threshold for Auger processes.

A semi-classical language is very useful for qualitatively understanding what stimulates the Auger process rate. The electron "likes" to take up the annihilation energy at the place where its kinetic energy, or its kinetic energy uncertainty, has a maximum value. For example, in the presence of an impurity, the Auger process takes place right at the center of the impurity, where the potential energy goes to minus infinity and the kinetic energy has a maximum.

In the case of Auger processes in NCs they take place right at the abrupt heterointerface, because of large uncertainty of the electron momentum. As a result electrons can get enough momentum at the interface. It is also directly seen in the quantum mechanical calculation of the matrix element $M$ (see Eq. 6) that Auger processes take place right at the nanocrystal surface. The abrupt surface significantly accelerates the Auger rate in NCs with large surface to volume ratio.

Figure 9 compares the result of the Auger rate calculations using Eq. 5 with experimental data indicated by black triangles [16]. The theoretical dependence of Auger autoionization rate $1/\tau_A$ on size (solid line) has a strong oscillatory character. The maxima of $1/\tau_A$ correspond to the NCs for which the electron quantum size level with angular momentum 1 ($P$ state) is very close the electron continuum or just becomes delocalized. The minimum of this dependence is due to the vanishing of the matrix element *M* that is caused by the quantum mechanical interference of the electron wave functions.

In real NCs these oscillations of $1/\tau_A$ are averaged out due to numerous reasons such as disordering of the glass matrix, dispersion of nanocrystal sizes and shapes, and fluctuations in the semiconductor glass band offset. The averaging of the theoretical size dependence in Fig. 9 gives the average rate:

$$\left\langle \frac{1}{\tau_A} \right\rangle \sim \frac{1}{a^5} \qquad (7)$$

that increases significantly with decreasing NC size. The experimental size dependence measured in CdS doped glasses (see Fig. 9) can be described as $\langle 1/\tau_A \rangle_{\exp} \sim 1/a^{4.5}$, and that is slightly weaker than the theoretical one.

Actually, the calculation shows that the rate of Auger autoionization depends strongly on band offset and the size dependence may be even stronger. For the average size dependence Eq. 5 gives:

$$\left\langle \frac{1}{\tau_A} \right\rangle \sim \frac{1}{a^\nu} \qquad (8)$$

where $5 < \nu < 7$, depending on the band offset. The theoretical dependence shown in Fig. 9 and Eq. 8 were calculated for the effective energy gap of the glass 7.02 eV obtained from direct photoionization data on CdS nanocrystals embedded in the glass [19].

The strong size dependence of an average Auger ionization rate described by Eqs. 7 and 8 also explains the shift of the PL line in porous-Si after the light induced degradation (see Fig. 10). In porous Si the size distribution of Si NCs is very broad. As a result the smallest NCs that before degradation contribute to the green part of the PL line are Auger ionized and do not contribute to the PL after degradation, while the largest Si crystals in the distribution stay neutral and contribute to the red edge of the PL line.

The oscillatory theoretical dependence of the rate of Auger ionization has not yet been observed. Still, this prediction shows that there are ways in principle to suppress considerably the rate of Auger processes by changing the shape of the nanocrystals and the value of the potential wall surrounding it by modifying the NC surface. This is a kind of Auger rate engineering.

**Random Telegraph Signal in the PL Intensity of a Single Quantum Dot**

Now, let us return to a single nanocrystal experiment that shows the random intermittence of the PL that resembles a random telegraph signal (RTS). Although solutions of the kinetic state-filling equations (Eq. 1) describe the time behavior of an ensemble of NCs very well [16], and also the time average behavior of a single dot, they do not describe a particular stochastic sequence of "on" and "off" periods occurring in a single NC.

The intermittence of the PL is a consequence of a particular sequence of events. Low intensity CW light creates electron-hole pairs which recombine radiatively emitting light ("on" period) until the quantum dot is ionized, either thermally or by Auger autoionization. The ejected electron (hole) is localized in a deep trap in the surrounding matrix, initiating the "off" period because nonradiative Auger processes in a charged crystal then quench all PL. The ejected electron (hole) returns to the NC and restores the "on" period of the PL.

This process is essentially a stochastic one. To simulate a particular sequence of "on" and "off" periods in a single quantum dot we construct a matrix of random transitions directly from Eq. 1. For example, a NC containing one electron-hole pair is transformed into one with two pairs in a small time interval $\Delta t$ with a probability $W_2 \Delta t$, into one with no pairs with a probability $\Delta t / \tau_1$, and remains unchanged with a probability of 1- $W_2 \Delta t - \Delta t / \tau_1$ (the total probability is equal to one). The transition probabilities for the other states are obtained similarly. To determine which transition occurs in $\Delta t$ we choose random numbers uniformly distributed in the interval [0,1] (a Monte Carlo procedure). We subdivide this interval into lengths equal to the appropriate probabilities described above. The subinterval in which a chosen random number falls determines which transition occurs. Thus we simulate the stochastic sequence of NC states as a function of time. Each step in this simulation depends only on the current NC state and not on the previous states.

Let us now discuss the parameters in Eq. 1 that play an important role in the RTS and determine the conditions under which it can be observed.

The radiative lifetime of a single electron-hole pair, $\tau_1$: In the case of strong size quantization, when the electron and hole motions are independent, the overlap integral of their wave functions does not depend on crystal size and is on the order of unity. [24] As a result, the radiative lifetime of a pair in a single dot, $\tau_1$, should be on the order of several nanoseconds for almost all semiconductors.[25] However, formation of the optically forbidden ground exciton state (dark exciton) in the NC greatly increases the decay time, at low temperature, to as much as several microseconds (depending on the NC size [26]). Filling of the higher optically active states at higher temperature lowers this time to several nanoseconds again and makes it a function of the NC radius.

The radiative lifetime of nanocrystals with two *e-h* pairs, $\tau_2$, is less than $\tau_1$ because each electron (hole) can recombine with either of two holes (electrons), and formation of an optically forbidden biexciton state is impossible. As a result this time is always on the order of nanoseconds.

We have discussed the Auger autoionization time, $\tau_A$, in a previous section. Generally, the NC can be Auger autoionized only if the band offset in the conduction or valence band is smaller than the NC energy gap. In CdS doped glasses, the measured conduction band offset satisfies this condition [19,20] and the ionized NC contains an unpaired hole. No measurements have been made of the band offset of CdSe NCs embedded in glass or polymers; however persistent [27] and dynamic [28] hole burning experiments performed on these samples clearly demonstrate the formation of an unpaired electron in the NCs. This electron occupies the first quantum size level of electrons and leads to the bleaching of corresponding optical transitions to this level. This is indirect proof of hole ejection from the NC. $\tau_A$ depends strongly on the NC radius, on the band offset at the boundary between the NC and the surrounding matrix (see Eqs. 7 and 8), and can be less than 0.1 ns in small NCs.

The time for the trapped electron (hole) to return to the NC, $\tau$, can be phenomenologically written $\tau=\tau_{ph}\exp(\Delta E/kT)$, where $\Delta E$ is the depth of the trap in the matrix, $T$ is the temperature, and $\tau_{ph}$ is a typical phonon scattering time in the matrix. At room temperature $\tau$ is about 10 min in semiconductor doped glasses [16]. For NCs embedded in polymers this time is much shorter (several seconds [10]), which probably reflects a smaller value of the trap depth there.

The single NC absorption cross section for excitation of the first electron-hole pair $\sigma_1(\omega)$, is on the order of $\sigma_1 \sim 10^{-16}$-$10^{-14}$cm$^{-2}$ and, far from the band edge, $E_g$, is proportional to the NC volume and to the electron and hole reduced density of states $\sigma_1(\omega) \sim \sqrt{\hbar\omega - E_g}$ in the parabolic band approximation [24] at the excitation energy $\hbar\omega$. The absorption cross section for excitation of a second electron-hole pair $\sigma_2(\omega)$ is approximately equal to $\sigma_1(\omega)$ far from the band edge, where the excitation energy is in the continuous spectrum. It can be much smaller than $\sigma_1$, however, for excitation near the band edge, where the discrete character of the excited electron-hole pair states plays an important role and excitation goes via phonon assisted transitions.

Figure 11 from Ref. [29] shows the results of a simulation of the RTS using parameters appropriate for nanosize CdSe crystals at excitation frequencies far from the band edge. We choose an intensity of excitation close to 0.08 kW/cm$^2$ in order to observe quite long periods of brightness. On average, at this intensity, the time interval between photon absorption is $1/W_1 = 5000$ ns. All relaxation processes are much faster than this. The probability for exciting a two electron-hole pair state is very low, but when it occurs the NC is either almost instantly ionized or returns to a zero pair state. We integrate the intensity of the luminescence signal over a time interval $\Delta t_m \sim 10$ms, as in the experiment of Ref. [10], i.e., we count the number of emitted photons in that time interval. One can see that the luminescence resembles a random telegraph signal (random pulses of equal magnitude) only in a narrow range of excitation intensity. Lowering the intensity by a factor of three greatly reduces the number of "off" periods seen, while a corresponding increase in the excitation intensity results in random luminescence impulses having random intensities. We also perform the calculation using a smaller value of $\tau_A$ in order to simulate the effect of a smaller effective band offset in bare NCs. The result, presented in panel d, clearly shows a shortening of the "on" period and a decrease of the PL quantum efficiency.

One can introduce the effective "on" and "off" states of quantum dots using the fact that the collecting time of the luminescence signal is much longer than all the dynamic processes taking place in the QD. The ratio of the probability of finding a QD in an "off" state to that for an "on" state determines the ratio of the lengths of the "off" and "on" average periods:

$$P_{off} / P_{on} = \tau / \tau_i \qquad (9)$$

where the nanocrystal ionization time $\tau_i$ in the case of Auger autionization is given by Eqs. 3 and 4. In the case of thermal autoionization of the QD with a small band offset, $\tau_i$ is given by the following rate of NC ionization [29]:

$$\frac{1}{\tau_i} = \frac{W_1 \tau_1}{1 + W_1 \tau_1} \frac{1}{\tau_T} \qquad (10)$$

where $1/\tau_T \sim \exp(-\Delta E_{off}/kT)$ is the probability for electron or hole to leave the nanocrystal thermally and where $\Delta E_{off}$ is the band offset between the NC and the matrix. In this case the ionization rate depends linearly on excitation intensity in contrast to the Auger ionization rate (that has square dependence on excitation intensity) and now depends strongly on temperature.

The observation of the random intermittence of the PL intensity in a single QD changes completely our understanding of the PL quantum efficiency, $\eta$, of an ensemble of nanocrystals. The PL quantum efficiency of the nanocrystal ensemble has been suggested to be determined by the ratio

$$\eta = (N - N_{nr})/N \qquad (11)$$

of the number of "good dots" ($N-N_{nr}$) – that do not contained nonradiative channels - to the total number of the dots (*N*). However, the fact that the luminescence is completely quenched in ionized crystals and that we can ionize these crystals by Auger or thermal autoionization gives us a completely different picture.

The PL quantum efficiency is determined rather by the ratio of the duration of the "on" period to the duration of the sum of "on" and "off" periods or by the ratio of the ionization time to the sum of ionization and neutralization times. In ideal QDs that do not contain nonradiative channels:

$$\eta = \tau_i/(\tau_i + \tau) \qquad (12)$$

As can be seen from Eq. 12 the quantum efficiency can be very low even at low excitation intensity if the time for the charge to return to the nanocrystal is long. The PL quantum efficiency even in an ensemble of ideal QDs depends on the temperature, size, and excitation intensity.

## Nonradiative Auger Relaxation in Nanocrystals with Several Electron-Hole Pairs

As we mentioned in a previous section, the nonradiative Auger recombination of two electron-hole pairs leads to crystal ionization only if the band offset in the conduction or valence band is smaller than the NC energy gap. However, even if that is the case, the electron or the hole of the electron-hole pair that receives the annihilation energy of the first one does not always leave the NC. The Auger annihilation does not always lead to the NC ionization. This type of nonradiative Auger process, when the recombination energy of one pair is transferred to the other one, is an essential part of the stimulated emission process. This is because the lasing from direct-band semiconductor NCs requires more than one electron-hole pair per crystal on average that provides an inverse population at the NC PL band edge.

The rate of multiparticle Auger recombination has been measured in an ensemble of CdSe nanocrystals in Ref. [30] at different excitation levels. The dynamics of the 2-, 3-, 4-pair states decay in a 2.3 nm radius CdSe nanocrystal is shown in Fig. 12. One can see that the carrier decays faster as the number of electron-hole pairs in the NC increases, as expected for Auger recombination. In the bulk semiconductor the effective decay rate in the Auger regime is inversely proportional to the square of the carrier concentration: $1/\tau_A(n) \sim n^2$. To use this expression in NCs one has to determine the concentration as $n = N_{e-h}/V$, where $N_{e-h}$ is the number of electron-hole pairs in the nanocrystal ($N_{e-h} > 1$) and $V$ is its volume. This bulk concentration dependence predicts for NCs with 4, 3 and 2 electron pairs a ratio of the Auger decay times $\tau_4 : \tau_3 : \tau_2 = 0.25:0.44:1$, respectively, that is very close to the ratio of 0.22:0.47:1 for the experimentally determined times (10, 21 and 45ps, respectively). The experimental results [30] show that the Auger decay rate for confined electron-hole pairs obeys the cubic concentration dependence: $dn_{e-h}/dt \sim (n_{e-h})^3$, the same as in bulk.

Figure 13 from Ref. [30] compares the dynamics of the two electron-hole pair decay for CdSe nanocrystals of different size. The time decay constant decreases from 363 to 6 ps when the nanocrystal radius, $a$, was reduced from 4.1 to 1.2 nm. The size dependence of $\tau_2$ has been shown in Ref. [30] to be approximated by a cubic dependence: $\tau_2(a) \sim a^3$.

There is no calculation of the decay time for the Auger nonradiative recombination of two electron-hole pairs. However, we can estimate this time using our calculations of the rate of Auger ionization (see Eq. 8). This can be written:

$$\frac{1}{\tau_2} \sim \left\langle \frac{1}{\tau_A} \right\rangle \rho(a) \sim \frac{1}{a^\nu} a^3 = \frac{1}{a^\beta} \qquad (13)$$

where $\rho(a)$ is the density of excited states in the nanocrystal that is proportional to the nanocrystal volume [24], and $2 < \beta < 4$. The difference of Auger nonradiative recombination from the Auger auto-ionization rate is connected with a finite density of the states where the Auger electron can be transfered. The resulting size dependence correlates well with the measured size dependence of the Auger nonradiative rate.

The nonradiative Auger process, where the recombination energy of one pair is transferred to the other pair, plays a fundamental role in the PL in NCs. The time of nonradiative Auger recombination decreases strongly with decreasing radius and becomes much smaller than the radiative recombination time. The saturation of the PL intensity in the NCs is one of the important consequences of the fast nonradiative recombination in NCs. Only NCs with one electron-hole pair contribute to the PL and the excitation of any additional pair does not increase the intensity of the PL. The effect is clearly seen in an ensemble of Si nanocrystals [31] where the time of electron-hole relaxation (on the order of microseconds at room temperature) is significantly longer than in CdSe nanocrystals. Figure 14 shows the excitation intensity dependence of the PL intensity from porous Si excited well above the PL line. One can see that the PL intensity at the red edge of the PL line saturates earlier than at its blue edge, which is connected with differences of the electron-hole radiative decay times in small and large NCs. The radiative decay time in Si NCs decreases with the NC due to an enhance of the oscillator strength of the phononless transitions. This is because the electron states in the X and $\Gamma$ points of the Brullouin zone are mixed in the small NCs. As a result the PL coming from the large NCs that contribute to the red edge of the PL line saturates earlier than the PL coming from the small NCs that contribute to the blue edge of the PL line.

## Auger-Like Thermalization in NCs

Not all of the Auger processes quench the PL in the NCs and decrease the PL quantum efficiency. For example, the Auger-like thermalization of carriers increases the PL quantum efficiency, because it is responsible for breaking a "phonon bottleneck" in small NCs. In small NCs the quantum size level spacing is much larger than the energy of an optical phonon. The direct transitions between levels have to be a multi-phonon process that has a very low probability. The standard phonon-assisted mechanism of carrier thermalization is then strongly suppressed: the effect is referred to as "phonon bottleneck." It was thought that this effect would make it impossible to use QDs for optical application.[32]

However, it was shown [33] that the rate of the Auger-like thermalization process is much faster than the radiative recombination time and it breaks the phonon bottleneck. This process is shown schematically in Fig. 15 from Ref. [34]. The spacing between the ground 1S and the first excited 1P electron levels is on the order of 200-300meV in small size NCs and is significantly larger than the optical phonon energy (20-30 meV) - this is the "phonon bottleneck." However, the electron-hole Coulomb interaction, although it is a small perturbation compared to the quantum size energy, becomes stronger in small NCs. In addition, hole level spacing are an order magnitude smaller than those of electrons, because of the higher hole effective mass and the degeneracy of the valence band. As a result the electron energy is transferred via the Coulomb interaction to the hole, which then relaxes rapidly through the almost continuous spectrum of valence band states. This process is called Auger-like thermalization.[1]

The straightforward calculation of the size dependence of the rate of the electron transition from the 1P level to the 1S level in CdSe nanocrystals shows that the Auger-like thermalization time is on the order of 2 ps and almost constant over the range of radii from 2 to 4 nm. This time is significantly shorter than the electron-hole pair radiative decay time in this material and thus inelastic electron-hole collisions remove the phonon bottleneck problem in CdSe nanocrystals.

Indeed, this mechanism of fast electron relaxation was experimentally confirmed recently. The authors of Ref. [35] and [36] have studied the dynamics of the electron thermalization from the 1P state to the 1S state. Both experiments show that removing the photoexcited hole from the nanocrystal to the nanocrystal surface significantly suppresses the rate of electron thermalization. Removing the hole from the NC decreases the probability of energy transfer from the electron to the hole and can be interpreted as suppression of this Auger-like thermalization mechanism.

---

[1] A similar type of the thermalization process when the electron energy is transferred to the hole via their Coulomb interaction has been studied in GaAs quantum wells. It is called inelastic scattering of the electron with heavy holes and is the next most efficient mechanism of the thermalization after thermalization via emission of the optical phonons.

## Concluding Remarks

The purpose of this review is to describe the role played by Auger processes in kinetics of the thermalization and interband relaxation of carriers. The Auger processes are considerably enhanced in NCs due to the abruptness of their interfaces and they affect all aspects of carrier relaxation and recombination: they strongly affect the PL quantum efficiency, and lead to PL blinking and reversible PL degradation. They require a significant shortening of the stimulated emission rate for lasing in order to overcome the fast nonradiative decay in the NC with two electron–hole pairs. The misunderstanding of their role explains why quantum dot NC laser realization was delayed for almost 10 years. They also shorten the rate of carrier relaxation and allow one to use semiconductor doped glasses as a holographic material and as material for X-ray sensors.

Although the theory of Auger processes in a NC qualitatively describes many aspects of these processes, a quantitative description of their rates has yet to be developed. This is connected with the lack of a self-consistent theory that describes the electron-hole envelope functions at the NC surface. In our calculations we have used the envelope functions obtained with the help of the standard boundary conditions that assume the separate continuity of the envelope function and the normal envelope velocity at the interface. This assumption has never been justified and is a special case of the more general boundary conditions that conserve the total current across the interface. The discontinuities in the envelope wave functions at the heterointerface is one of the most important consequences of general boundary conditions.[37] They could strongly affect the calculation of the Auger processes rate, because they take place right at the nanocrystal surface and are affected by the surface property. The theoretical solution of this problem is important, because it may open the way to controlling the Auger processes rate through surface modifications.

## Acknowledgements

I would like to thank my co-authors: V. A. Kharchenko and M. Rosen, who worked with me on the theory of Auger processes in nanocrystals. I also would like to thank my long–term collaborators: A. I. Ekimov, M. G. Bawendi, D. Norris, M. Nirmal, P. Alivisatos, as well as A. Nozik and L. Brus, whose challenging discussions stimulated most of these theoretical studies. I also thank M. Nirmal, V. Klimov and J. Cook for helping me with the figures. This work was supported by the Office of Naval Research.

## Figure Captions

Fig. 1. The time dependence of the room temperature PL intensity of a single 2.1 nm radius CdSe NC under steady state excitation of $I_{exc}$=0.7kW/cm$^2$ and a sampling interval

of 20ms. The upper and lower panels show the dependence for a "bare" NC and an overcoated one with a shell thickness of ~ 7 monolayers of ZnS respectfully (from Ref. [10]).

Fig. 2. Comparison of the time dependence of the room temperature PL of a single 2.1 nm radius CdSe NC overcoated by a 4 monolayer thickness ZnS shell. The upper and lower panels show the dependence measured at excitation intensity $I_{exc}$=0.52kW/cm$^2$ and $I_{exc}$=0.1.32kW/cm$^2$, respectively. The average on/off times at the two intensities are: $<\tau_{on}>$ ~0.97s, $<\tau_{off}>$ ~0.44s at $I_{exc}$=0.52kW/cm$^2$ and $<\tau_{on}>$ ~0.32s, $<\tau_{off}>$ ~0.43s at $I_{exc}$=1.32 kW/cm$^2$, respectively.

Fig. 3. The time dependence of the PL intensity $I_{lum}$ of 2.2nm radius CdS NCs at different intensity of excitation: curves 1, 2 and 3 corresponds to $I_{exc}$, 0.45 $I_{exc}$, and 0.2$I_{exc}$, respectively. The insert shows the PL spectra of this sample.

Fig. 4. The dependence of ionization time $\tau_i$ on absorbed power for different size CdS NCs. The straight lines: 1, 2, 3, and 4 approximate these dependencies for NCs with average radius: 1.3, 1.6, 2.2 and 3.9 nm, correspondingly. Filled and open circles show experimental data measured at the excitation by the light with the wavelength of 400 nm and 476 nm respectfully.

Fig. 5. The thermostimulated luminescence (TSL) curves measured in CdS doped glass (solid lines) and in undoped glass ( dotted line). Both samples were irradiated at 80K for 1 h.

Fig. 6. Spectral dependence of the efficiency of ionization $S_{TSL}$ for the glass doped by 3 nm radius CdS NCs (curve 1) and for undoped glass (curve 2).

Fig. 7. A scheme presenting various types of relaxation processes in a NC under steady state excitation by visible light. The ionized NC that quenches the PL can be created by visible light as result of Auger or thermal autoionization.

Fig. 8. The dependence of ionization time $\tau_i$ on the average CdS NC size at fixed power of absorbed light $P$=1 mW (from Ref. [16]).

Fig. 9. The size dependence of the Auger autoionization time of a NC with two electron-hole pairs, $\tau_A$. The crosses show experimental data. The solid line is the theoretical dependence calculated with use of Eq. 5 (from Ref. [16]).

Fig. 10. The effect of optical degradation by a 20 min. illumination by a 20W/cm$^2$ 488 nm line of a CW Ar$^+$ laser pump beam on the intensity and spectral position of the PL line in porous Si (from Ref. [22]).

Fig. 11. The number of phonons emitted in the simulation in 10 ms interval by a single NC quantum dot. We use $\tau_1$=4 ns, $\tau_2$=1 ns $\tau$=0.8 s, $\sigma_1 = \sigma_2 = 1.0 \times 10^{-15}$ cm$^2$, $\hbar\omega_{ex}$=2.5 eV (from Ref. [29]).

Fig. 12. The decay dynamics of the 1-, 2-, 3-, and 4- electron-hole pair states in the 2.3 nm radius CdSe NCs. Experimental data extracted from transition absorption measurements are shown by symbols. The solid lines are their fit to a single exponential decay dependence (from Ref. [30]).

Fig.13. The time decay of the two electron-hole pair states measured in the NC quantum dots with radii 1.2, 1.7, 2.3, 2.8, and 4.1 nm (symbols are shown in the insert). The experimental data are fit by a single exponential dependencies (from Ref. [30]).

Fig. 14. Dependence of the room temperature PL intensity in porous-Si on the excitation intensity for three different detection energies: 2.07 eV ( squares), 1.77 eV ( circles) and 1.55 eV ( triangles). The excitation energy is 2.54 eV. The insert shows the PL spectra.

Fig. 15. A scheme presenting an Auger-like thermalization process. The electron energy is transferred via the Coulomb interaction to the hole, which then relaxes rapidly through the almost continuous spectrum of valence band states.

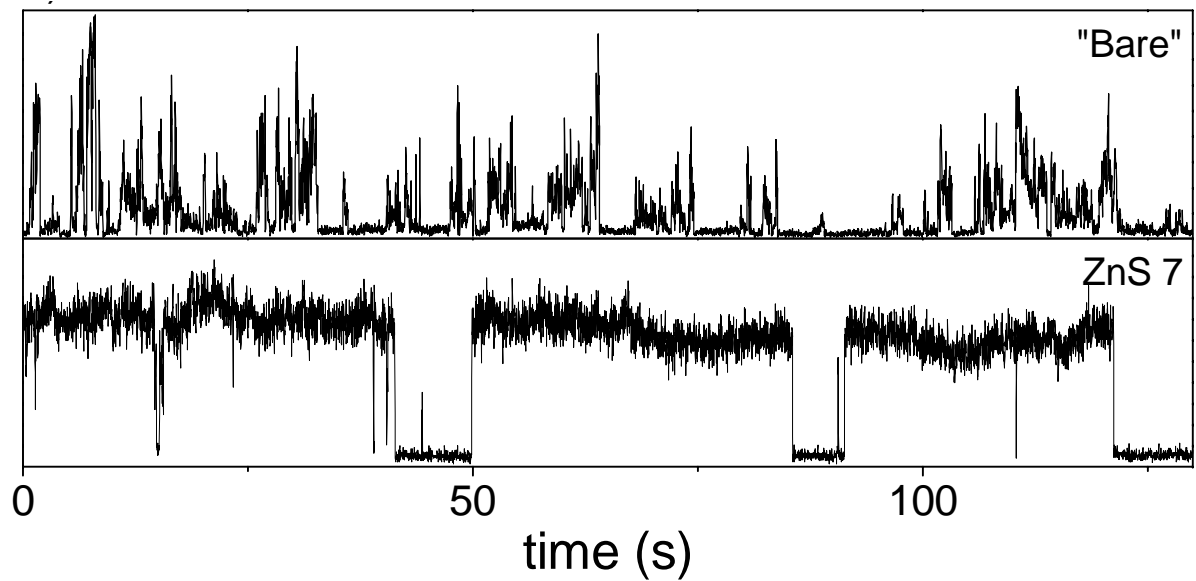

Fig.1 Al. L. Efros, "Auger…."

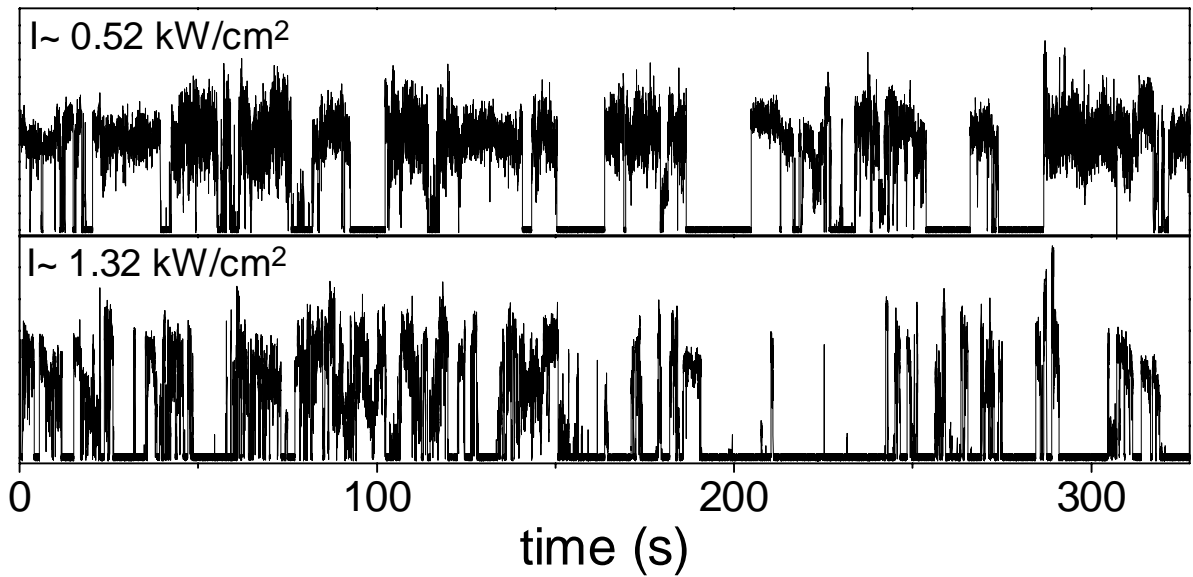

Fig.2 Al. L. Efros, "Auger…."

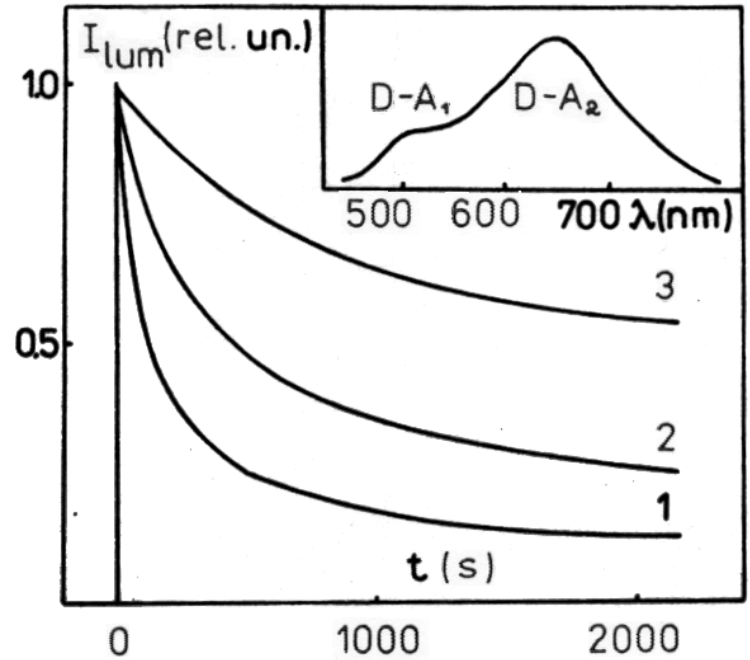

Fig.3 Al. L. Efros, "Auger…."

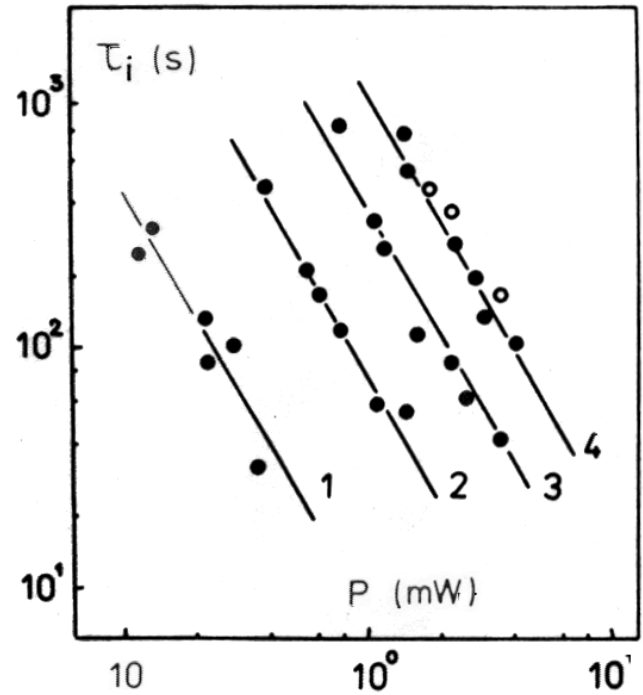

Fig.4 Al. L. Efros, "Auger…."

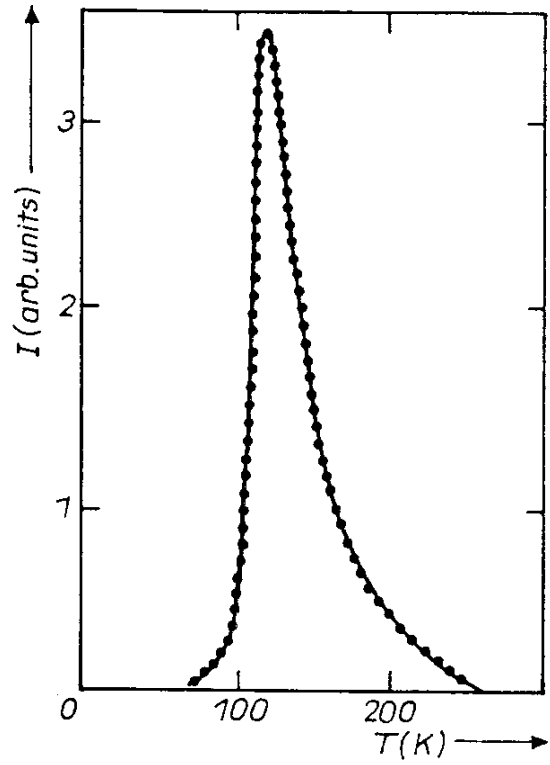

Fig.5 Al. L. Efros, "Auger…."

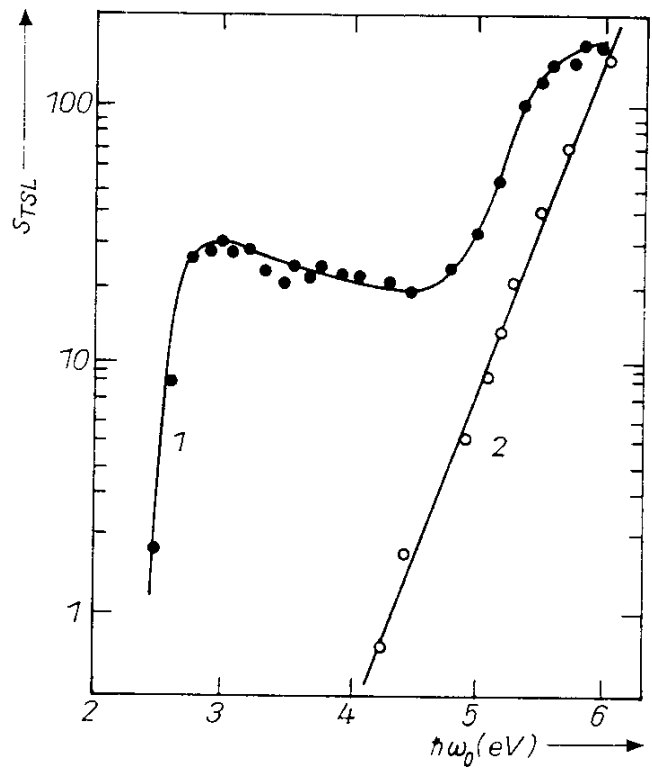

Fig.6 Al. L. Efros, "Auger…."

Fig.7 Al. L. Efros, "Auger…."

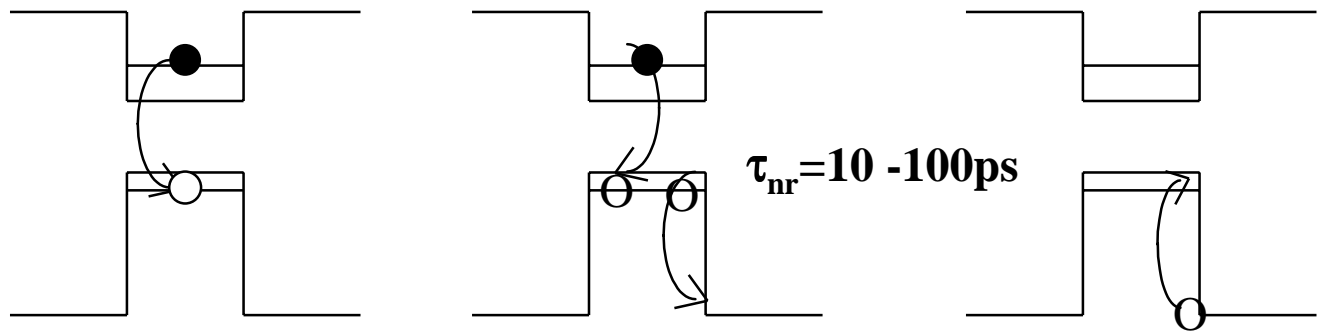

Photoluminescence  Auger quenching

$\tau_{nr}$=10 -100ps

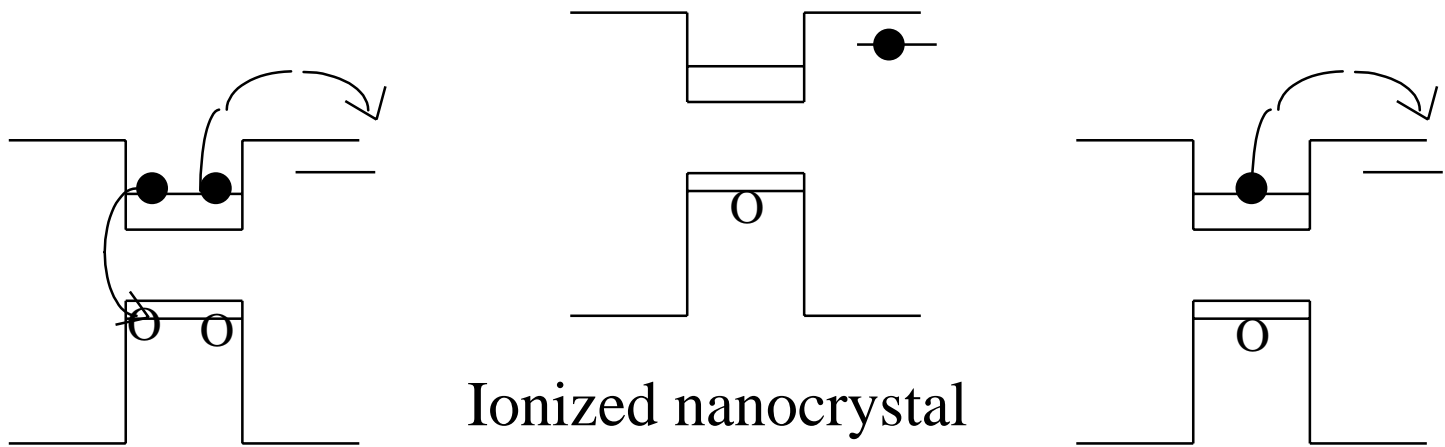

Ionized nanocrystal

Auger autoionization  Thermal autoionization

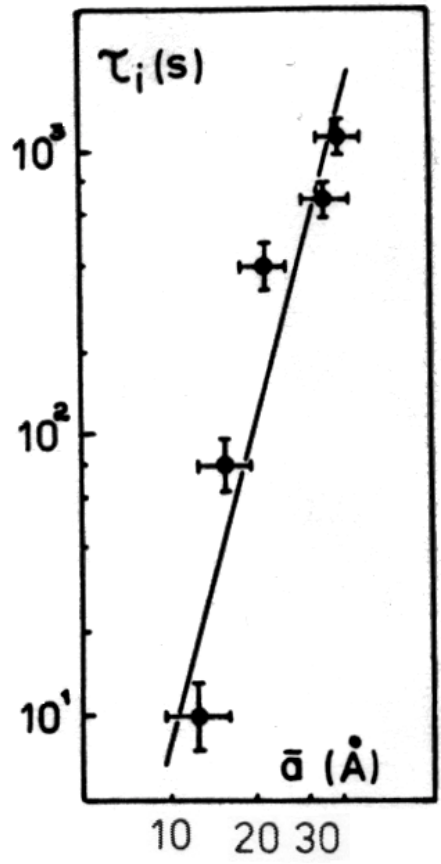

Fig.8 Al. L. Efros, "Auger…."

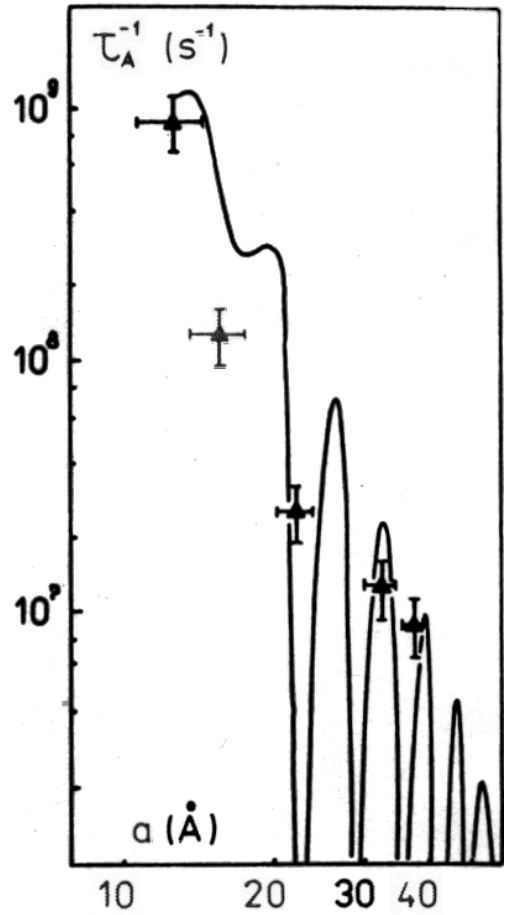

Fig.9 Al. L. Efros, "Auger…."

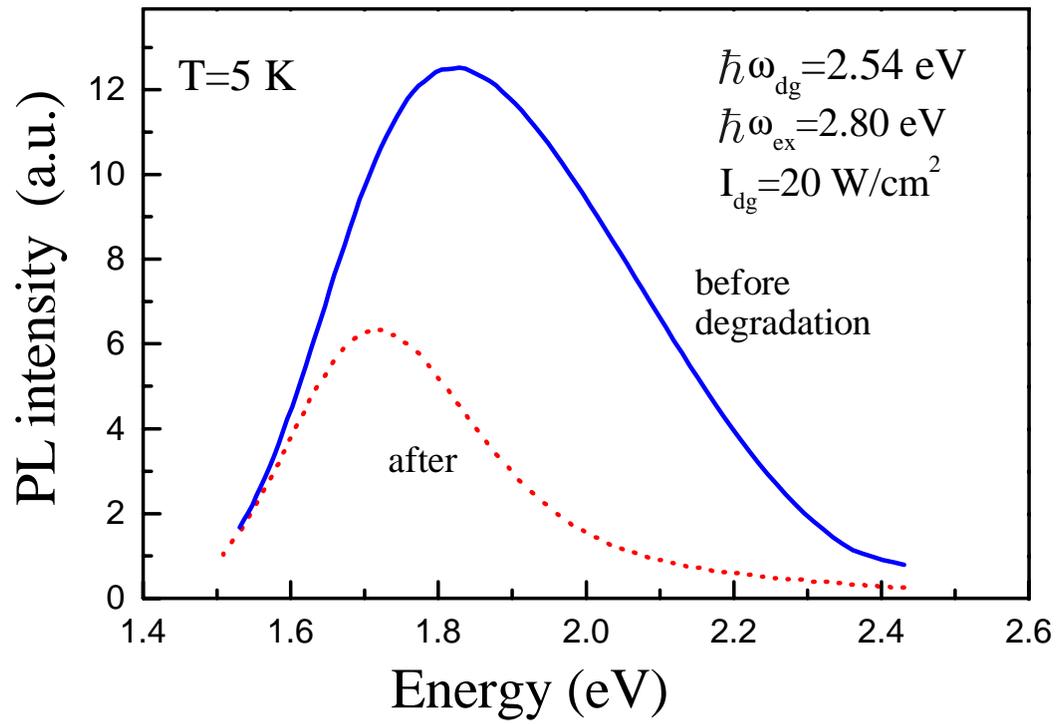

Fig.10 Al. L. Efros, "Auger…."

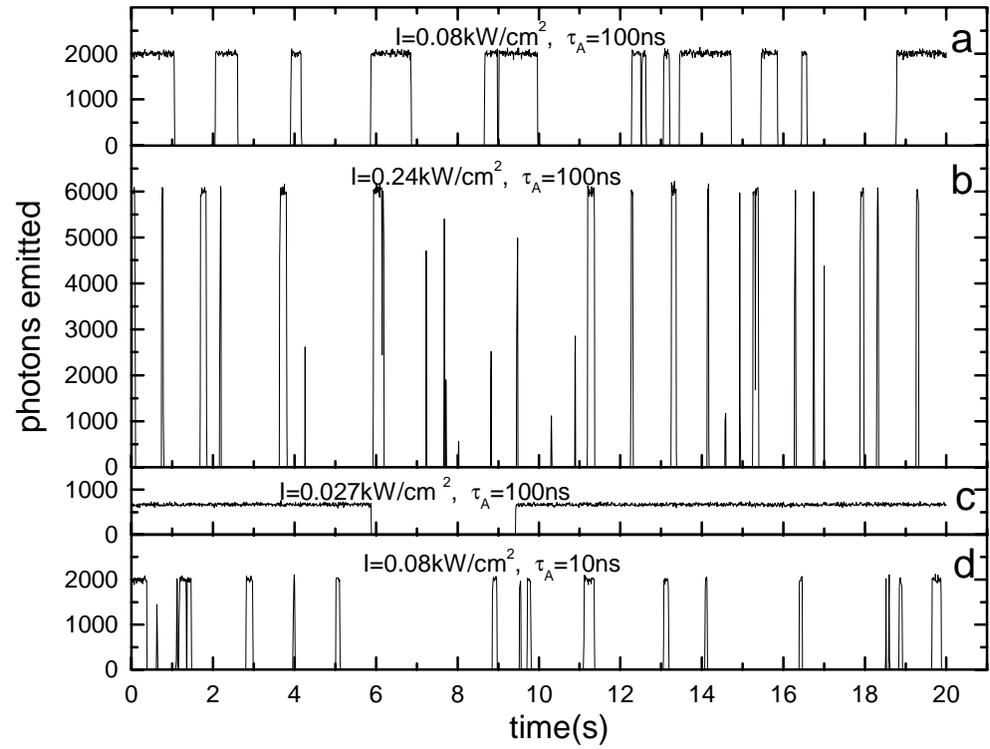

Fig.11 Al. L. Efros, "Auger…."

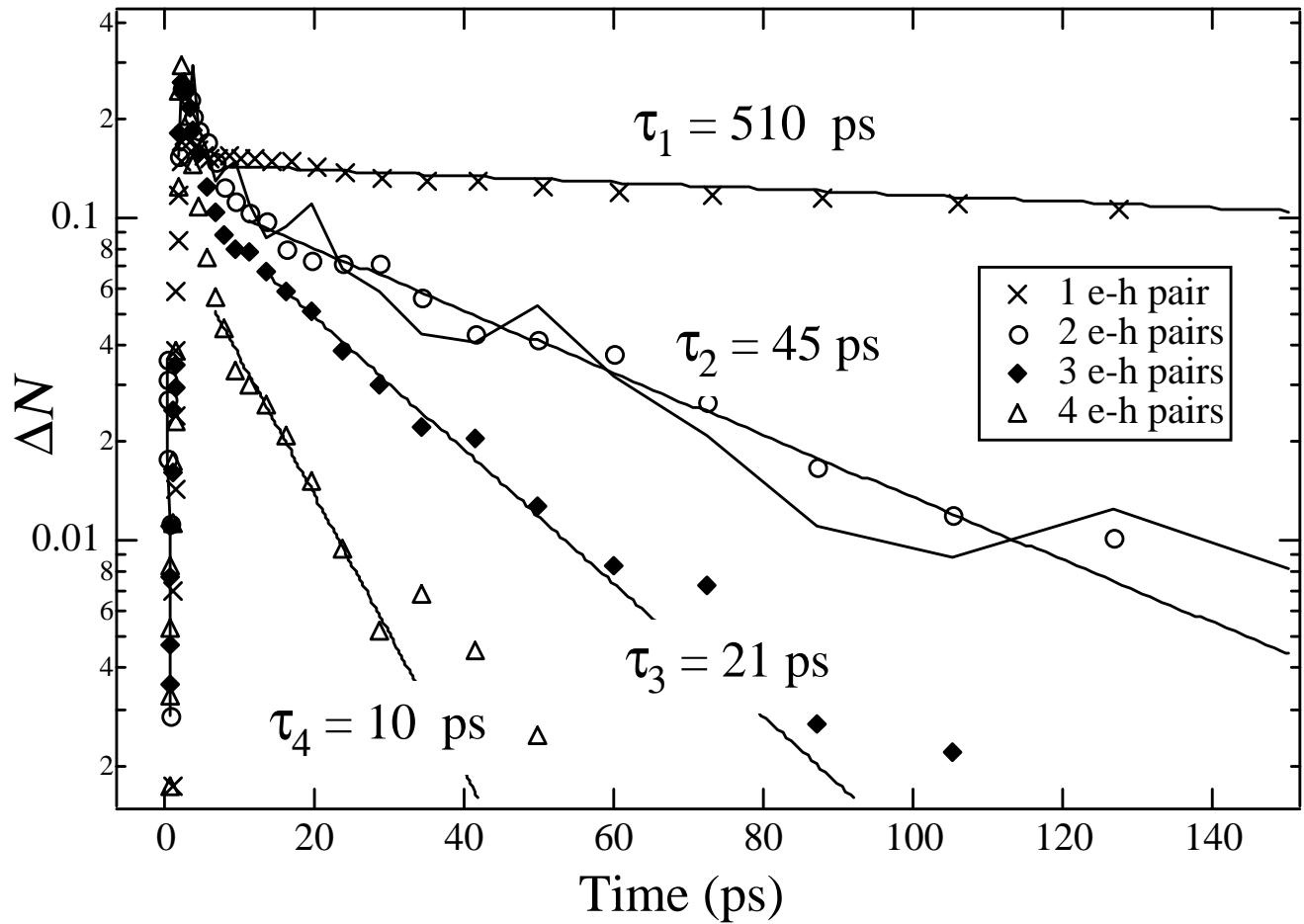

Fig.12 Al. L. Efros, "Auger…."

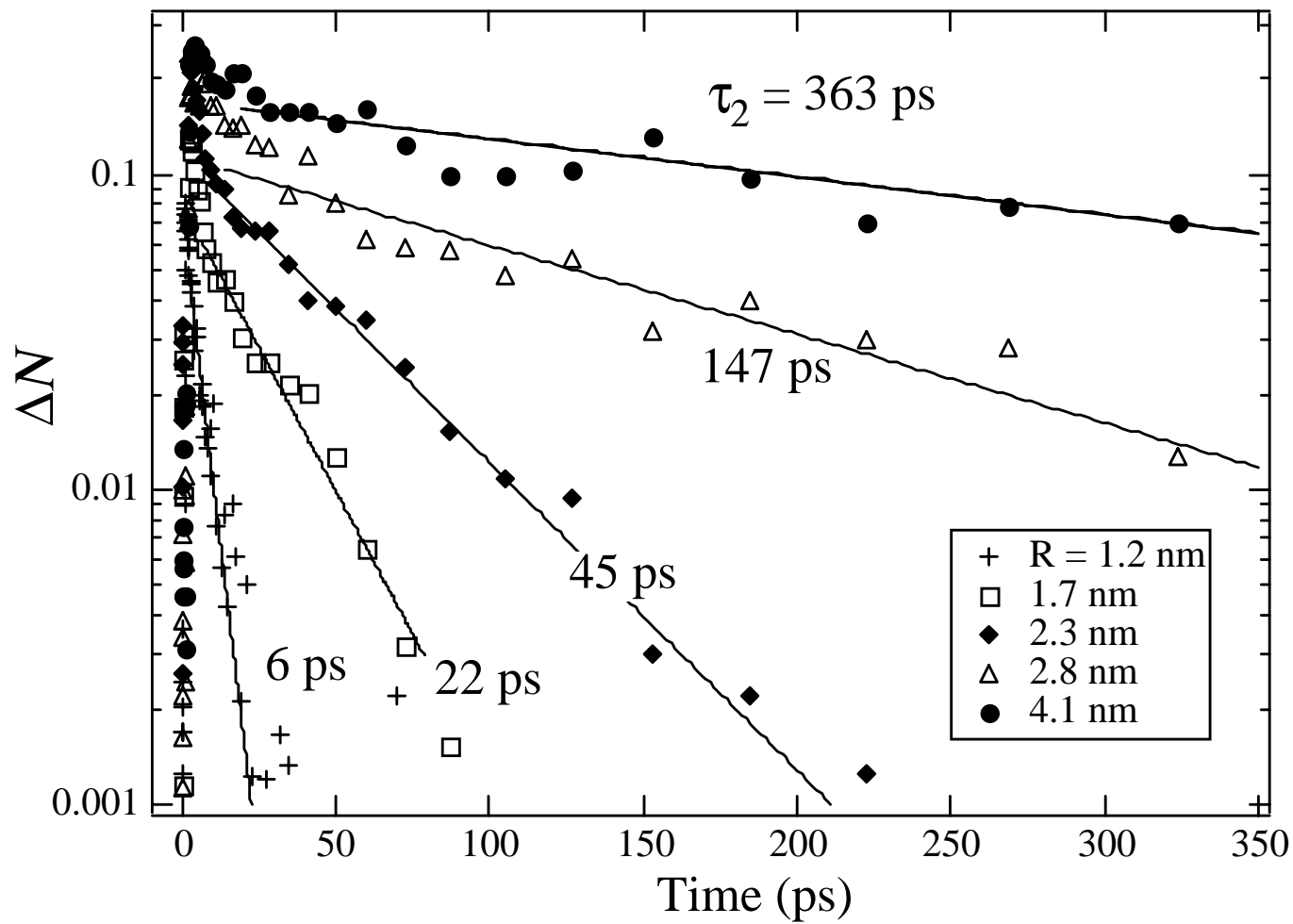

Fig.13. Al. L. Efros, "Auger…."

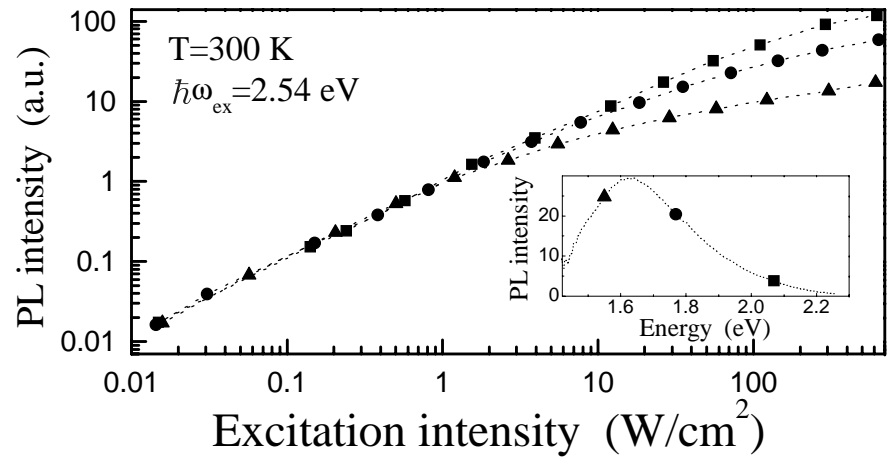

Fig.14 Al. L. Efros, "Auger…."

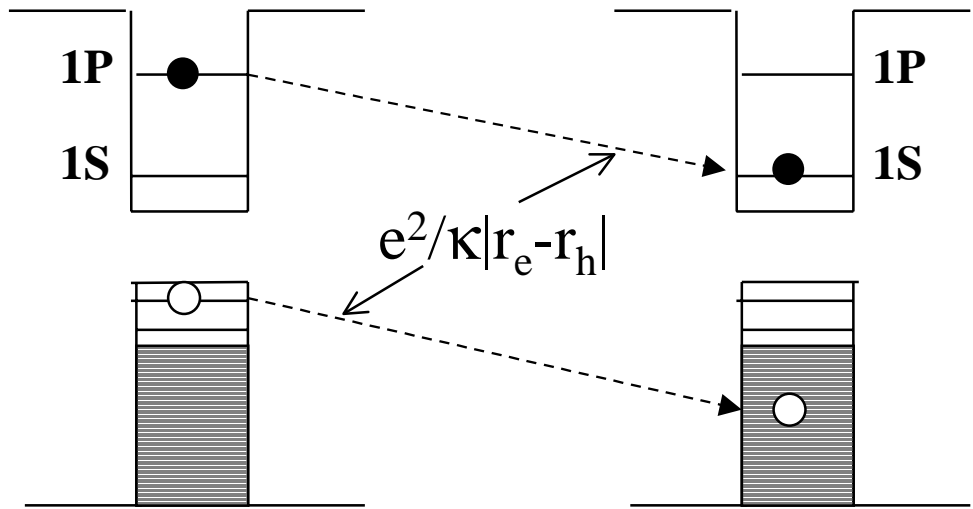

Fig.15. Al. L. Efros, "Auger…."